\documentclass[aps,prb,twocolumn,showpacs,superscriptaddress,groupedaddress]{revtex4-1}  

\usepackage{graphicx}
\usepackage{dcolumn}
\usepackage{bm}
\usepackage{amssymb}
\usepackage{psfrag}
\usepackage{amsmath}
\usepackage{epsfig}
\usepackage{subfigure}
\usepackage{grffile}
\usepackage{color}
\usepackage{morefloats}
\usepackage{color}
\usepackage{xr, xspace}

\graphicspath{{../Figures/}}

\newcommand{\beq}    {\begin{equation}}
\newcommand{\enq}    {\end{equation}}
\newcommand{\ceq}[1] {(\ref{#1})}
\newcommand{\kk}{{\bf k}}
\newcommand{\rr}{{\bf r}}
\newcommand{\qq}{{\bf q}}
\newcommand{\mm}{{\bf m}}
\newcommand{\cmm}{{\bf M}}
\newcommand{\sss}{{\bf s}}

\renewcommand\Re{\operatorname{Re}}

\newcommand{\sbte}     {${\rm Sb_2Te_3}$\xspace}
\newcommand{\bise}     {${\rm Bi_2Se_3}$\xspace}
\newcommand{\bite}     {${\rm Bi_2Te_3}$\xspace}
\newcommand{\df}     {\equiv}
\begin{document}

\title{Giant Edelstein effect in Topological-Insulator--Graphene heterostructures}
\author{M. Rodriguez-Vega}
\affiliation{Department of Physics, College of William and Mary, Williamsburg, VA 23187, USA}
\affiliation{Department of Physics, Indiana University, Bloomington, Indiana 47405,
	USA}
\author{G. Schwiete}            
\affiliation{Department of Physics and Astronomy, Center for Materials for Information Technology (MINT), The University of Alabama, Alabama 35487, USA}
\author{J. Sinova}
\affiliation{ 
             Institut f\"ur Physik, Johannes Gutenberg Universit¨at Mainz, 55128 Mainz, Germany.}
\affiliation{Institute of Physics, Academy of Sciences of the Czech Republic, Cukrovarnicka 10, 162 53 Praha 6 Czech Republic}
\author{E. Rossi}
\affiliation{Department of Physics, College of William and Mary, Williamsburg, VA 23187, USA}

\date{\today}
\begin{abstract}
The control of a ferromagnet's magnetization via only electric currents
requires the efficient generation of current-driven spin-torques.
In magnetic structures based on topological insulators (TIs)
current-induced spin-orbit torques can be generated. 
Here we show that the addition of graphene, or bilayer graphene, to a TI-based magnetic
structure greatly enhances the current-induced spin density accumulation and  significantly reduces
the amount of power dissipated.
We find that 
this enhancement can be as high as  a factor of 100, giving rise to a giant Edelstein effect.
Such a large enhancement is due to the high mobility of graphene (bilayer graphene) and to
the fact that the graphene (bilayer graphene) sheet very effectively screens charge impurities,
the dominant source of disorder in topological insulators.
Our results show that the integration of graphene in spintronics devices can greatly enhance
their performance and functionalities.

\end{abstract}

\maketitle

\section{Introduction}
The ability to generate and control spin currents in condensed matter systems has
led to several discoveries of great fundamental and technological interest ~\cite{zutic2004, sinova2015}.
In recent years the discovery of whole new classes of materials with strong spin-orbit coupling,
such as topological insulators (TIs)~\cite{hasan2010,qi2011},
has allowed the realization of novel basic spin-based phenomena~\cite{Sinova2012,brataas2012b,Jungwirth2012b,Bauer2012b}.

In a system with spin-orbit coupling (SOC), a charge current ($I$) can induce a spin-Hall effect (SHE)~\cite{sinova2015}
i.e. a pure spin-polarized current. 
A companion effect to the SHE, also arising from the SOC,
is the inverse spin-galvanic effect (ISGE),
where a current induces a non-equilibrium uniform spin accumulation~\cite{dyakonov1971,edelstein1990,dyakonovbook,sinova2015}. 
In a magnetic system this current-driven spin accumulation results in 
a spin-orbit torque (SOT)  acting on the magnetization (\cmm),
and therefore can be exploited to realize current-driven magnetization dynamics.
The SOT, ${\boldsymbol{\tau}}_{SO}$, can
be either an (anti-)damping torque~\cite{Kurebayashi2014, sinova2015}, i.e. have the same functional
form as the Gilbert damping term, or {\em field-like}~\cite{sinova2015}, i.e. 
have the form
$\boldsymbol{\tau}_{SO}=\gamma{\bf B}_{SO}\times{\bf M}$, where ${\bf B}_{SO}$ 
is an effective spin-orbit field, and $\gamma$ is the gyromagnetic ratio.
The presence of a current-driven SOT on the surface of TIs
has been predicted~\cite{garate2010, yokoyama2010, sakai2014, fischer2013,ndiaye2015}
and it has been recently measured
in TI-ferromagnet bilayers~\cite{mellnik2014b} and magnetically doped TIs~\cite{fan2016}.
\begin{figure}[!t]
	\includegraphics[width=8.5cm]{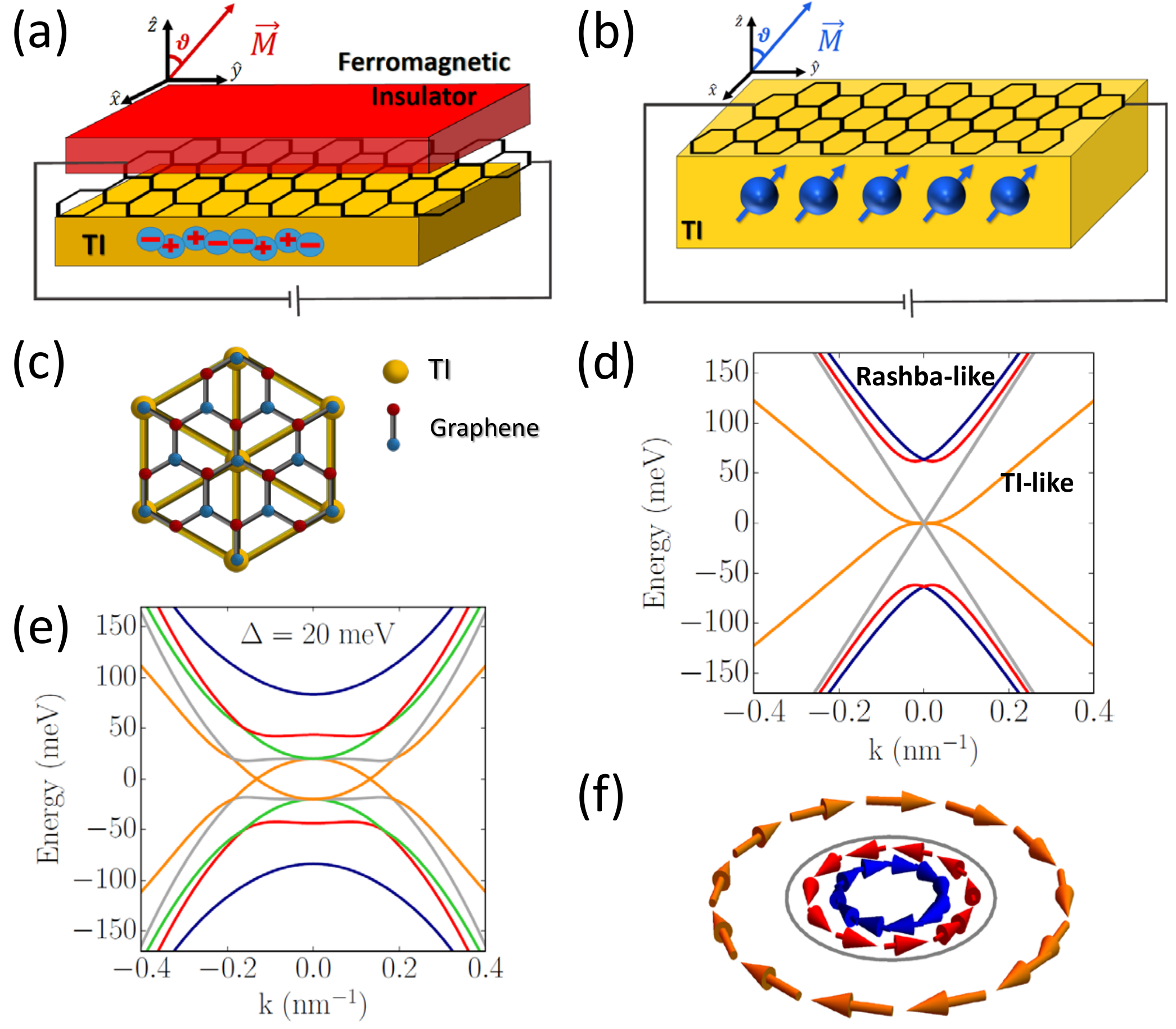}
	\caption{Sketch of a TI-graphene-FM, (a), and of a magnetically doped TI-graphene, (b), heterostructure.
		In (a) the random charges are shown. In (b) the spheres represent
		magnetic dopant, the random charges are not shown explicitly.
		(c) Atoms' arrangement for the commensurate stacking considered.
		(d) Bands for TI-SLG for  $\Delta=0$, $\delta\mu=0$. 
		(e) Bands for TI-BLG for $\Delta=20$~meV and $\delta\mu=0$.
		(f) Spin texture on the Fermi surface formed by the bands shown in (d) for $\epsilon_F=100$~meV.
	}
	\label{fig:bands}
	%\vspace{-0.65 cm}
\end{figure}

The two-dimensional nature of graphene (SLG) and bilayer graphene (BLG)~\cite{novoselov2005,zhang2005,castroneto2009}
and the fact that their room-temperature mobilities are higher than in any other known material~\cite{dassarma2011}
make them extremely interesting for transport phenomena.
However, the SOC in graphene is extremely small and as a consequence graphene
alone is not very interesting for spintronics applications, except as a spin-conductor.
%The proximity effect of graphene on insulating ferromagnets substrates has been
%theoretically studied in the context of *** .
Several methods have been proposed to induce larger SOC in graphene \cite{leejeongsu2016}.
Recent experiments on TI-graphene heterostructures seem to demonstrate the injection of spin-polarized current
from a TI into graphene~\cite{vaklinova2016,tian2016}.

In this work we show that the combination of a particular class of three-dimensional (3D) TIs and graphene allows the realization
of devices in which a charge current induces a spin density accumulation that can be up
to a factor 100 larger than in any previous system, 
i.e. a {\em giant Edelstein effect}.
% 
%
%
%For most of the experimentally relevant conditions considered 
We find that for most of the experimentally relevant conditions considered 
the SOT in TI-graphene vdW heterostructures should be higher than 
% by at least a factor 10
%larger than 
the already very large values observed in TI-Ferromagnet bilayers~\cite{mellnik2014b}
and magnetically doped TIs~\cite{fan2016}.
In Ref.~\onlinecite{mellnik2014b}, for $I=7.7~{\rm mA}$, a $B_{SO}=3\times 10^{-2}$~mT 
was measured, in Ref.~\onlinecite{fan2016}, for $I=~4\mu{\rm A}$, a $B_{SO}=80$~mT was measured~\cite{note2016}.
%ER-NN---
%{\bf XX}
Assuming that our work is able to capture the key elements affecting the SOT
in TI-graphene systems we find that in these systems the SOT could be ten times larger than the 
%already
%very large 
values found in Ref.~\onlinecite{mellnik2014b,fan2016}.
%
%Starting from these values our results show
%
%
%
%that the $B_{SO}$ created by the ISGE effect
%can be higher than $0.3$~mT in TI-BLG-Ferromagnet devices, Fig~\ref{fig:bands}~(a), 
%and $800$~mT in vdW heterostructures formed by a magnetically doped TI and BLG, Fig~\ref{fig:bands}~(b).
%
We also find that TI-SLG and TI-BLG systems have conductivities much
% that are a factor of 10 and 50, respectively,
higher than TI surfaces and would therefore allow the realization of 
spintronics effects with dramatically lower dissipation than in TIs alone.

The rest of the paper is organized as follows: 
in Sec. \ref{sec:theoretical_framework} we introduce the
effective model for the TI-graphene heterostructure,
describe the treatment of disorder, and outline
the calculation of the current-induced spin density response
function; in Sec. \ref{sec:numerical} we present our
results; finally, in Sec. \ref{sec:conclusions} we present our conclusions. 

\section{Theoretical framework}
\label{sec:theoretical_framework}
In  vdW heterostructures~\cite{haigh2012}, the different layers are held together by vdW forces.
This fact greatly enhances the type of heterostructures
that can be created given that the stacking 
is not fixed by the chemistry of the elements forming the heterostructure.
With $a=2.46 \mathring{A}$ being the lattice constant of graphene, and $a_{TI}$ the lattice constant of the 111 surface 
of a TI in the tetradymite family, 
we have $a_{TI}/a=\sqrt{3}(1+\delta)$, where $\delta<1\%$ 
for \sbte, $\delta=-3\%$ for \bise, and $\delta=3\%$ for \bite.
As a consequence, graphene and the 111 surface of \sbte, \bise, \bite,
to very good approximation, can be arranged in a $\sqrt{3}\times\sqrt{3}$
commensurate pattern ~\cite{jin2013,falko2013,jzhang2014}.
When the stacking is commensurate the hybridization between the graphene's and the TI's surface states 
is maximized.
%when graphene and the TI's surface
%are stacked in a commensurate configuration.
%
This property of graphene, combined with its high mobility, its intrinsic two-dimensional nature, and its ability at finite dopings
to effectively screen the dominant source of disorder in TIs,
make graphene the ideal material to consider for creating a TI heterostructure with a very large Edelstein effect.

TI-graphene heterostructures can be formed via mechanical 
transfer~\cite{steinberg2015b,bian2016b,tian2016}. 
As a consequence, the stacking pattern and the shift are 
fixed by the exfoliation-deposition process and can be 
controlled~\cite{ponomarenko2013}. Density functional theory (DFT) results 
show that the binding energy between graphene and the TI’s surface depends 
only very weakly on the rigid shift~\cite{jin2013,spataru2014,lee2015,rajput2016}. 
Among the commensurate configurations with free energy close to the minimum, 
as obtained from DFT calculations~\cite{jin2013}, we consider the stacking configuration
shown in Fig.~\ref{fig:bands}~(c). For this configuration, we expect the Edelstein effect to 
be the smallest because the graphene bands split into Rashba-like bands (see Figs.~\ref{fig:bands}~(d),~(e)), 
that give an Edelstein effect with opposite sign to the one given by TI-like bands~\cite{mellnik2014b}. 
Therefore, to be conservative, in the remainder we consider both the commensurate case 
for which the Edelstein effect is expected to be the weakest (i.e., the case in the 
graphene sublattice symmetry is broken) and the extreme case in which the tunneling 
strength between the TI and graphene is set to zero.

At low energies, the Hamiltonian for the system can be written as $H=\sum_\kk\psi^\dagger_\kk H_\kk\psi_\kk$
where $\psi^\dagger_\kk$ ($\psi_\kk$) is the creation (annihilation) spinor for a fermionic
excitation with momentum $\kk$, and
\beq
\rm H_\kk  =
 \begin{pmatrix}
  \hat{H}^{G, K}_\kk & 0                      & \hat{T}^{\dagger} \\
  0                     & \hat{H}^{G, K'}_\kk & \hat{T}^{\dagger} \\
  \hat{T}              & \hat{T}              & \hat{H}_\kk^{TI} 
 \end{pmatrix}\;,
  \hspace{0.5cm}
  \hat{T} =
  \begin{pmatrix}
  t  & 0 & 0 & 0   \\
  0  & 0 & t & 0   \\
 \end{pmatrix}\;,
 \label{eqn:full_hamiltonian}
\enq
where $\hat{H}^{G, K}_\kk$ ($\hat{H}^{G, K'}_\kk=[\hat{H}^{G, K}_\kk]^*$) is the 
Hamiltonian describing graphene's low energy states around the $K$ ($K'$) of the Brillouin zone.
For SLG $\hat{H}^{G, K}_\kk=\hat{H}^{SLG,K}$ and for BLG $\hat{H}^{G, K}_\kk=\hat{H}^{BLG,K}$.
$\hat{H}^{TI}_\kk$ is the Hamiltonian describing the TI's surface states,
and $\hat{T}$ is the matrix describing spin- and momentum- conserving tunneling processes between the graphene layer and the TI's surface~\cite{jzhang2014},
$t$ being the tunneling strength.
%
%ER-NN02
The TI's bulk states are assumed to be gapped.
This condition is realized, for example, in novel  ternary or quaternary tetradymites, such as
${\rm Bi_2Te_2Se}$ and ${\rm Bi_{2-x}Sb_xTe_{3-y}Se_y}$, for which it has been shown experimentally that the
bulk currents have been completely eliminated \cite{ren2010, xiong2012, arakane2012, xia2013, segawa2012, xu2014d, durand2016, xu2016}.
For SLG we have
$\hat{H}^{SLG, K}_\kk=\hbar v_g k\sigma_0\left[ \cos (\phi_\kk) \tau_x +\sin (\phi_\kk) \tau_y \right]- \mu_g$,
where $v_g\approx 10^6$~m/s is the graphene's Fermi velocity,
$k=|\kk|$, $\phi_\kk=\arctan(k_y/k_x)$, $\sigma_i$, $\tau_i$ are the
Pauli matrices in spin and sublattice space respectively,
and $\mu_g$ is the chemical potential.
For BLG we have
$\hat{H}^{BLG, K}_\kk = \hbar^2 k^2/(2m^*) \sigma_0\left[ \cos (2\phi_\kk) \tau_x + \sin (2\phi_\kk) \tau_y \right]- \mu_g$,
where $m^*\approx0.035 m_e$ is the effective mass.
For the TI's surface  states, we have 
$
\hat{H}^{TI}_\kk  =  \hbar v_{TI} \left( k_y \sigma_x - k_x \sigma_y \right) -\mu_{\rm TI} \sigma_0\,,
$
where $v_{TI}\approx v_g/2$ and $\mu_{\rm TI}$ is the chemical potential on the TI's surface .
In the remainder the Fermi energy $\epsilon_F$ is measured from the neutrality point of the SLG (or the BLG) 
and $\delta \mu \df \mu_{TI}-\mu_g$.

In a magnetically doped TI, below the Curie temperature,
the low energy Hamiltonian for the TI-graphene's quasiparticles, Eq.~\ceq{eqn:full_hamiltonian},
has an additional term, $H_{ex}$,
describing the exchange interaction between the quasiparticles and the magnetization $\cmm$.
$H_{ex}=\Delta\int_{\Omega}\hat\mm\cdot\hat\sss\;d\rr/\Omega$, where
$\Delta$ is the strength of the 
exchange interaction,
$\hat{\bf m}\df \cmm/|\cmm|$,
$\hat{\bf s}\df {\bf s}/{|\bf s|}$ with 
$\sss$ the TI-graphene's spin density operator, and $\Omega$ is the 2D area of the sample.  
%
%For a TI-graphene-ferromagnet heterostructure the ferromagnet (FM) will also cause simply the
%addition of the term $H_{ex}$ to the Hamiltonian for the quasiparticles,
%Eq.~\ceq{eqn:full_hamiltonian}, as long as the FM is an insulator, as is the case for the recently
%studied ${\rm Bi_2Se_3-EuS}$ systems~\cite{Lee2016,Katmis2016}.
%In the remainder, for TI-graphene-FM heterostructures we assume the FM to be an insulator.
%
For a TI-graphene-ferromagnet heterostructure the ferromagnet (FM) will also cause simply the
addition of the term $H_{ex}$ to the Hamiltonian for the quasiparticles,
Eq.~\ceq{eqn:full_hamiltonian}, as long as the FM is an insulator, 
and is placed on graphene, or bilayer graphene,
via mechanical exfoliation, likely with a large twist angle
to minimize hybiridization. Recent experiments have
studied ${\rm Bi_2Se_3-EuS}$ systems~\cite{Lee2016,Katmis2016}.
In the remainder, for TI-graphene-FM heterostructures we assume the FM to be an insulator.

To maximize the effect of the current-induced spin accumulation 
on the dynamics of the magnetization, it is ideal to have $\cmm$
perpendicular to the TI's surface. 
This is the 
case for magnetically doped TIs such as ${\rm Cr_{2x}(Bi_{0.5}Sb_{0.5-x})_2Te_3}$~\cite{fan2016}.
For TI-graphene-FM trilayers this can be achieved, for example, by using
a thin film of ${\rm BaFe_{12}O_{19}}$,
a magnetic insulator with high $T_c$ and large perpendicular anisotropy~\cite{yang2014}.
%

%
%
%ER--NN
%{\bf XX}
%Using Eq.~\ceq{eqn:full_hamiltonian} we obtain the bands of TI-SLG and TI-BLG at {\em low energies}.
By comparing the bands for TI-SLG, at low-energies, obtained from Eq.~\ceq{eqn:full_hamiltonian}, Fig.~\ref{fig:bands}~(d),
with the ones obtained using DFT~\cite{jin2013,lee2015,rajput2016} we obtain that the effective value of $t$ is $\sim 45$~meV. 
For this reason, most of the results that we show in the remainder were obtained assuming $t=45$~meV.
%In the supplementary material (SM) we also consider different values of $t$, see Fig.~4 in the SM. 
Fig.~\ref{fig:bands}~(d) clearly shows that, in general, the hybridization of the graphene's and TI's states 
preserves a TI-like band and induces the formation of spin-splitted Rashba bands.
The TI and Rashba nature of the bands can clearly be evinced from the winding of the spins,
as shown in Fig.~\ref{fig:bands}~(f).
The same qualitative features can be observed in Fig.~\ref{fig:bands}~(e) that shows
the low energy bands of a TI-BLG system with $\Delta=20$~meV, and $\delta\mu=0$.
%---------
%
%
%Fig.~\ref{fig:bands}~(e) shows 
%the band structure for a TI-BLG with $\Delta=20$~meV, and $\delta\mu=0$.
%In general, the hybridization of the graphene's and TI's states 
%
%gives rise to Rashba bands that are
%then splitted by the exchange term.
% due to the finite magnetization $\cmm = M_0\hat z$.
%The Rashba-like nature of some of the bands is evident from Fig.~\ref{fig:bands}~(f) that
%shows the spin texture on the Fermi surfaces obtained from the bands shown in panel (d)
%for $\epsilon_F=100$~meV.
%
%Fig.~\ref{fig:bands}~(f) also shows that on the larger Fermi surface, to very good approximation,
%the spins wind as on the Fermi surface of the isolated TI. 
%
%
%

In the remainder, we restrict our analysis to the case in which
$\epsilon_F$ is such that the system is metallic.
In this case contributions to the Edelstein effect from interband-transitions~\cite{sinitsyin2007} can be neglected
and the SOT is primarily field-like.
%ER-NN----
%{\bf XX}
%We also assume that the phase coherence length is much smaller than
%the system size so that quantum-interference effects are negligible.
For most of the conditions of interest, quantum interference effects 
can be assumed to be negligible due to dephasing effects
at finite temperatures and the large dimensionless conductance 
of the system.
%----------
%
%
%The SOT can be obtained by calculating 
%the current-induced spin density response function $\chi^{s_i J_j}(\qq,\omega)$ that,
%within the linear response regime, is equal to the spin-current
%correlation function.
The SOT can be obtained by calculating 
the current-induced spin density accumulation
$\delta s^i = \chi^{s_i J_j}(\qq,\omega) E^j$, where
$E^j$ is the electic field applied in the $j$-direction and
the spin density response function $\chi^{s_i J_j}(\qq,\omega)$,
within the linear response regime, is equal to the spin-current
correlation function.
%
%Considering that the SOT is given by ${\bf B_{so}}\times {\bf M}$,
%where  ${\bf B_{so}}=\delta {\bf s}$ is the effective spin-orbit field due to 
%the Edelstein effect, it is straightforward to see that the angular 
%dependence of the spin-orbit torque is simply $\sin(\vartheta)$, 
%where $\vartheta$ is the angle formed by the magnetization and the 
%TI's surface (see Fig. (\ref{fig:bands}a)). In the remainder, we will only
%consider the case $\vartheta=0$.
%
Considering that the SOT is given by ${\bf B_{so}}\times {\bf M}$,
where  ${\bf B_{so}}=\delta {\bf s}$ is the effective spin-orbit field due to 
the Edelstein effect, and that the response function depends weakly
on the gap $\Delta$ induced by ${\bf M}$ (as we demonstrate later),
the angular dependence of the torque is mainly geometrical. Without loss of generality, 
we can assume the external current to be in the $y$ direction so that ${\bf \delta s} \parallel \hat x$ and therefore 
${\boldsymbol{\tau}}_{SO} \approx |{\bf M}| \delta s^x \cos \vartheta  \left[ -\hat y +\hat z \tan \vartheta \sin \phi \right]$,
where $\vartheta$ is the angle formed by the magnetization and the TI's surface 
(see Fig. (\ref{fig:bands}a)) and $\phi$ is the angle with respect to $\hat x$ in the TI surface plane.

\begin{figure}[!t]
	\centering
	\includegraphics[width=8.5cm]{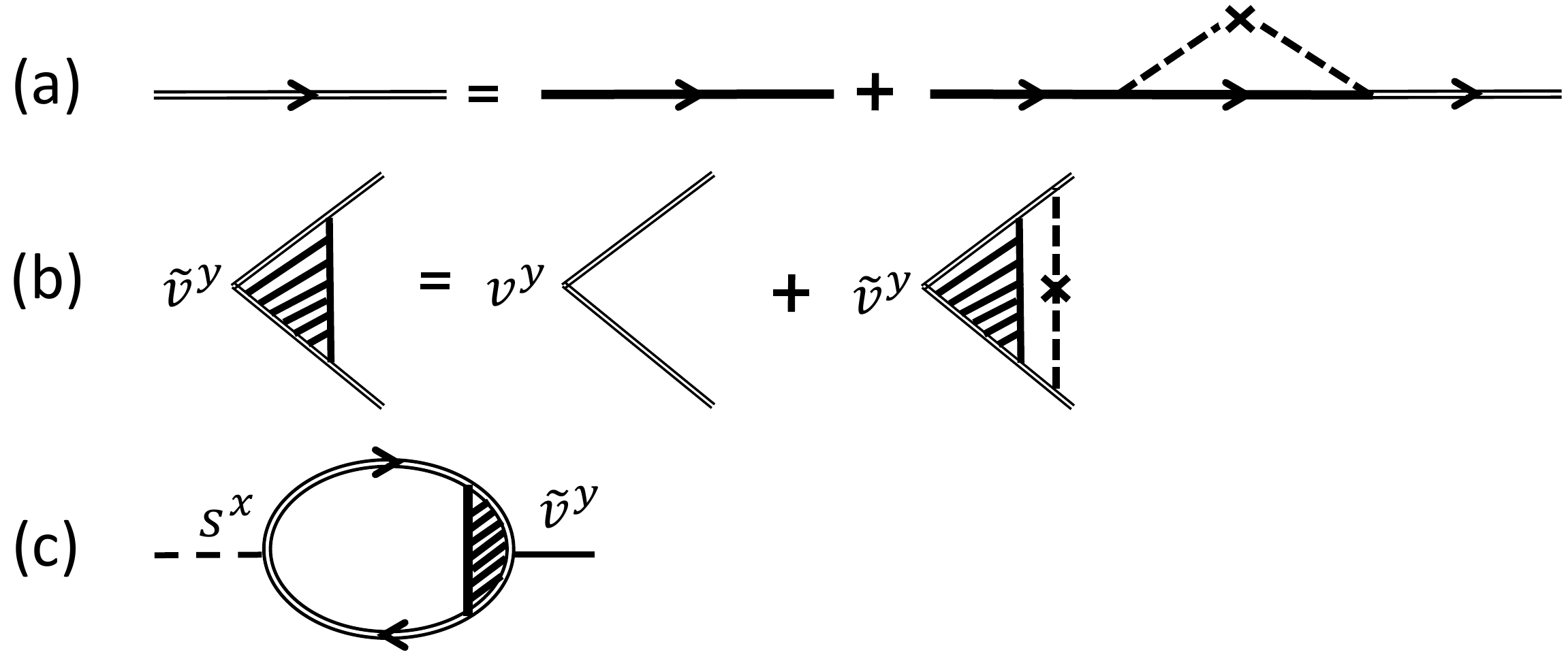}
	\caption{Diagrams used to calculate the charge conductivity and the spin-current response.}
	\label{fig:diagrams_1}
\end{figure}

The unavoidable presence of disorder induces a broadening of the quasiparticle states, and vertex corrections 
that are captured by the diagrams shown in Fig. \ref{fig:diagrams_1}.
%
%ER-NN-----
%{\bf XX}
In TIs charge impurities appear to be the dominant source of disorder~\cite{kim2012XXX}
and so it is expected that they will also be in TI-graphene heterostructures.
%----------
%
We therefore model the disorder as a random potential created by
an effective 2D distribution of uncorrelated charge impurities with zero net charge placed at an effective distance $d$
below the TI's surface.
%ER-NN-----
%{\bf XX}
Direct imaging experiments~\cite{beidenkopf2011} suggest $d\approx 1$~nm, consistent
with transport results~\cite{kim2012XXX, li2012b}.
%----------

%
In momentum space, the bare potential $v(q)$ created on the TI's surface by a single charge impurity is
$v(q)=2\pi e^2 e^{-q d}/(\kappa q)$ where $\kappa = (\kappa_{TI} +\kappa_0)/2 $ is the average dielectric constant
with $\kappa_{TI}\approx 100$~ ~\cite{culcer2010,butch2010,beidenkopf2011,kim2012XXX,li2012b} 
the dielectric constant for the TI and $\kappa_0=1$ the dielectric constant of vacuum~\cite{note2016b}.
The screened disorder potential is $v(q)/\epsilon(q)$ where $\epsilon(q)$
%is the 2D random phase approximation 
is the dielectric function~\cite{hwang2007,dassarma2011,triola2012}.
To obtain the current-driven SOT in the dc limit, and for temperatures $T$ much lower than the Fermi temperature $T_F$,
to very good approximation we can assume $\epsilon(q)\approx 1+v_c(q)\nu(\epsilon_F)$,
where $v_c(q)=2\pi e^2/(\kappa q)$ and $\nu(\epsilon_F)$ is the density of states at the Fermi energy.

The lifetime $\tau_{0a}(\kk)$ of a quasiparticle in band $a$ with momentum $\kk$ is given by
\beq
\frac{\hbar}{\tau_{0a}(\kk)} = 2\pi \sum_{a' \qq} n_{\rm imp} \left|  \frac{v(q)}{\epsilon(q)} \right| ^2 \lvert \langle a'  \kk+\qq | a \kk  \rangle \rvert ^2 \delta(\epsilon_{a, \kk}-\epsilon_{a', \kk+\qq}),
\label{eq:singleparttime}
\enq
where $n_{\rm imp}$  is the impurity density and $| a \kk  \rangle$ is the Bloch state with momentum $\kk$ and band index $a$.
In the remainder, we set $n_{\rm imp}=10^{12}{\rm cm}^{-2}$. \cite{kim2012XXX}
The transport time $\tau_{ta}(\kk)$, that renormalizes the expectation value of the velocity operator,  is obtained by introducing the factor $[1-\kk\cdot(\kk+\qq)]$ under the sum on the right hand side of Eq.~\ceq{eq:singleparttime},
and in general differs from the lifetime $\tau_{0a}(\kk)$~\cite{dassarma1985,nomura2007,rossi2008,rossi2009,lu2016}. 

For a charge current in the $y$ direction the non equilibrium spin density
is polarized in the $x$ direction.
Due to the rotational symmetry of the system
we have
$\chi^{s_x J_y} = -\chi^{s_y J_x}$ and 
$\chi^{s_x J_x} = \chi^{s_y J_y}$. 
Without loss of generality we can assume the current to be in the $y$ direction.
% and set $i=x$ and $j=y$.
%
Within the linear response regime, taking into account the presence of disorder,
the response function $\chi^{s_x J_y}$ of the system can be obtained by calculating the diagrams
shown in Fig.~(\ref{fig:diagrams_1}). The diagram in Fig.~(\ref{fig:diagrams_1}a) represents the
equation for the self-energy in the first Born approximation, where the double line represents 
the disorder-dressed electrons' Green's function, the single line the electron's Green's function 
for the clean system, and the dashed lines scattering events off the impurities. 
The diagram in Fig.~(\ref{fig:diagrams_1}b) corresponds to the
equation for the renormalized velocity vertex, $\tilde{v}_y$, at the ladder level approximation.
In the long wavelength, dc, limit we have
\begin{align} 
\chi^{s_x J_y} & \approx   \frac{e}{2 \pi \Omega} \Re \sum_{\kk, a}  s^x_{aa}(\kk)\tilde{v}^y_{aa}(\kk) G^A_{\kk a}G^R_{\kk a}\;,
\end{align}
where $s^i_{aa}(\kk)\df \langle a\kk|s_i|a\kk\rangle$ is the expectation value of the {\em i}-th component of the spin density operator,
$\tilde{v}^i_{aa}(\kk) = (\tau_{ta}/\tau_{0a})_\kk v^i_{aa}(\kk)$ with $v^i_{aa}(\kk)\df \langle a\kk|v_i|a\kk\rangle$
the expectation value of the 
{\em i}-th component of the velocity operator ${\bf v}=\hbar^{-1} \partial H_{\kk}/\partial \kk$,
and $G^{R/A}_{\kk a}=(\epsilon_F-\epsilon_{\kk a}\pm i\hbar/2\tau_{0a}(\kk))^{-1}$
are the retarded and advanced Green's functions, respectively, for electrons with momentum $\kk$ and band index $a$.

\section{Results}
\label{sec:numerical}
In this section, we present our results for the transport properties and
current-induced spin density accumulation of TI-graphene heterostructures.

\begin{figure}[!t]
	\includegraphics[width=8.5cm]{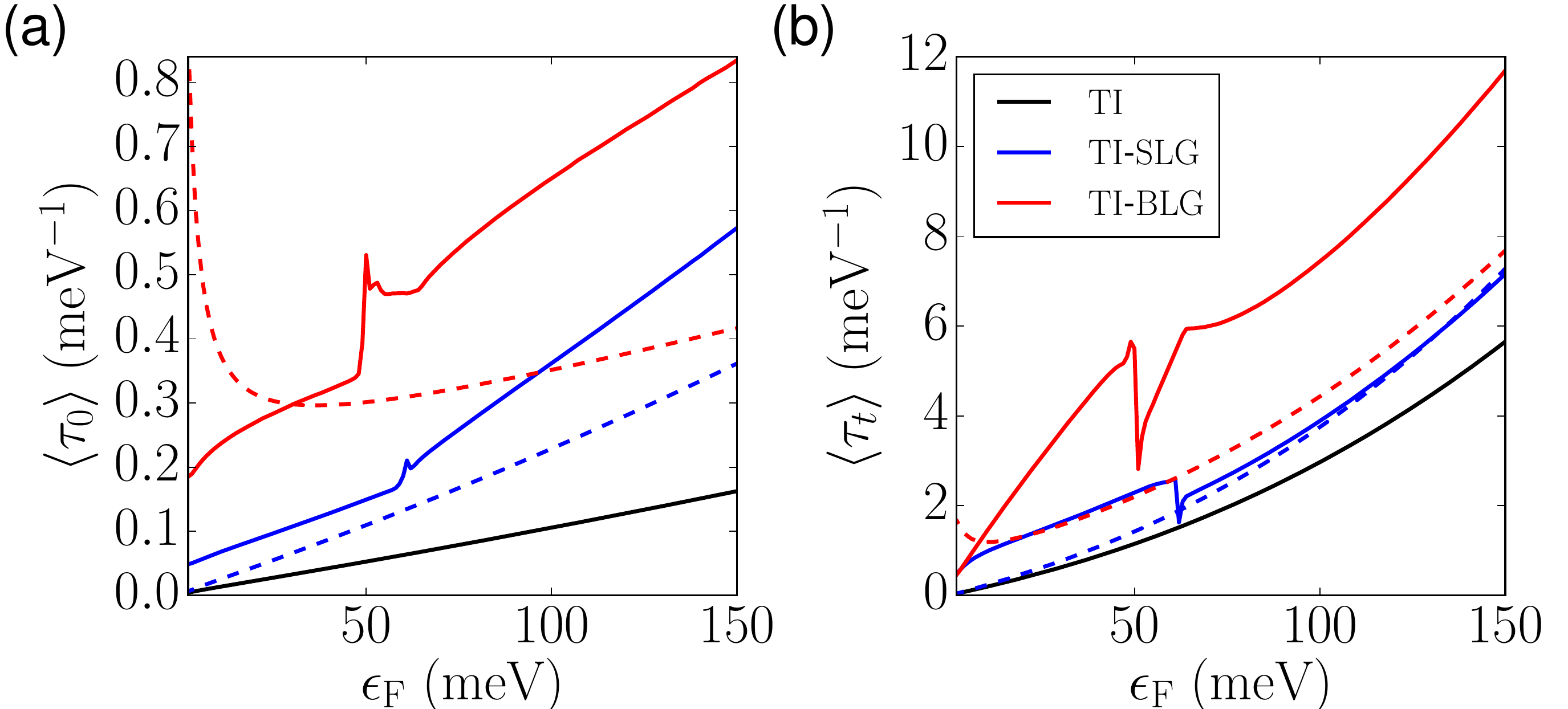}
	\caption{(a) $\langle\tau_{0}(\epsilon_F)\rangle$, and (b) $\langle\tau_t(\epsilon_F)\rangle$
		for $\Delta=0$~meV, $\delta \mu=0$~meV, and $n_{\rm imp}=10^{12}$ cm$^{-2}$. The solid lines
		correspond to $t=45$~meV while dashed lines to the limit $t=0$~meV.
	}
	\label{fig:gamma_comparison}
\end{figure}

We define the average $\tau_0$, $\tau_t$ as
$\langle\tau_{0(t)}(\epsilon_F)\rangle\df \sum_{\kk a}\tau_{0a(ta)}(\kk)\delta(\epsilon_F-\epsilon_{\kk a})/
\sum_{\kk a}\delta(\epsilon_F-\epsilon_{\kk a})$. Figs.~\ref{fig:gamma_comparison}~(a), (b) show  $\langle\tau_{0}(\epsilon_F)\rangle$, and $\langle\tau_t(\epsilon_F)\rangle$,
respectively, for a TI's surface, a TI-SLG  heterostructure, and a  TI-BLG heterostructure, with $\Delta=0$~meV.
We see that the presence of a graphene layer strongly increases both $\langle\tau_{0}(\epsilon_F)\rangle$, and $\langle\tau_t(\epsilon_F)\rangle$,
and that such increase is dramatic for the case when the layer is BLG. 
%
%ER-NN-------
%{\bf XX}
$\langle\tau_{0}(\epsilon_F)\rangle$, and $\langle\tau_t(\epsilon_F)\rangle$ are larger in BLG-TI than TI-SLG
because, especially at low energies, BLG has a larger density of states than SLG causing $\varepsilon(q)$, that enters in
the denominator in Eq.~\ceq{eq:singleparttime}, and therefore $\langle\tau_{0}(\epsilon_F)\rangle$, and $\langle\tau_t(\epsilon_F)\rangle$ 
to be larger in BLG than in SLG.
%
%------------
%
%ER-NN2
Notice that $\tau_0$, $\tau_t$ increase after adding a graphene layer even in the limit when $t=0$ as shown by the dashed lines
in Fig.~\ref{fig:gamma_comparison}. This is due to the fact that the graphene layer screens the dominant source of disorder
in the TI even when $t=0$.
Changes in $\Delta$ have only minor quantitative effects as long as $\Delta<(t,\epsilon_F)$.

\begin{figure}[!t]
	\includegraphics[width=8.5cm]{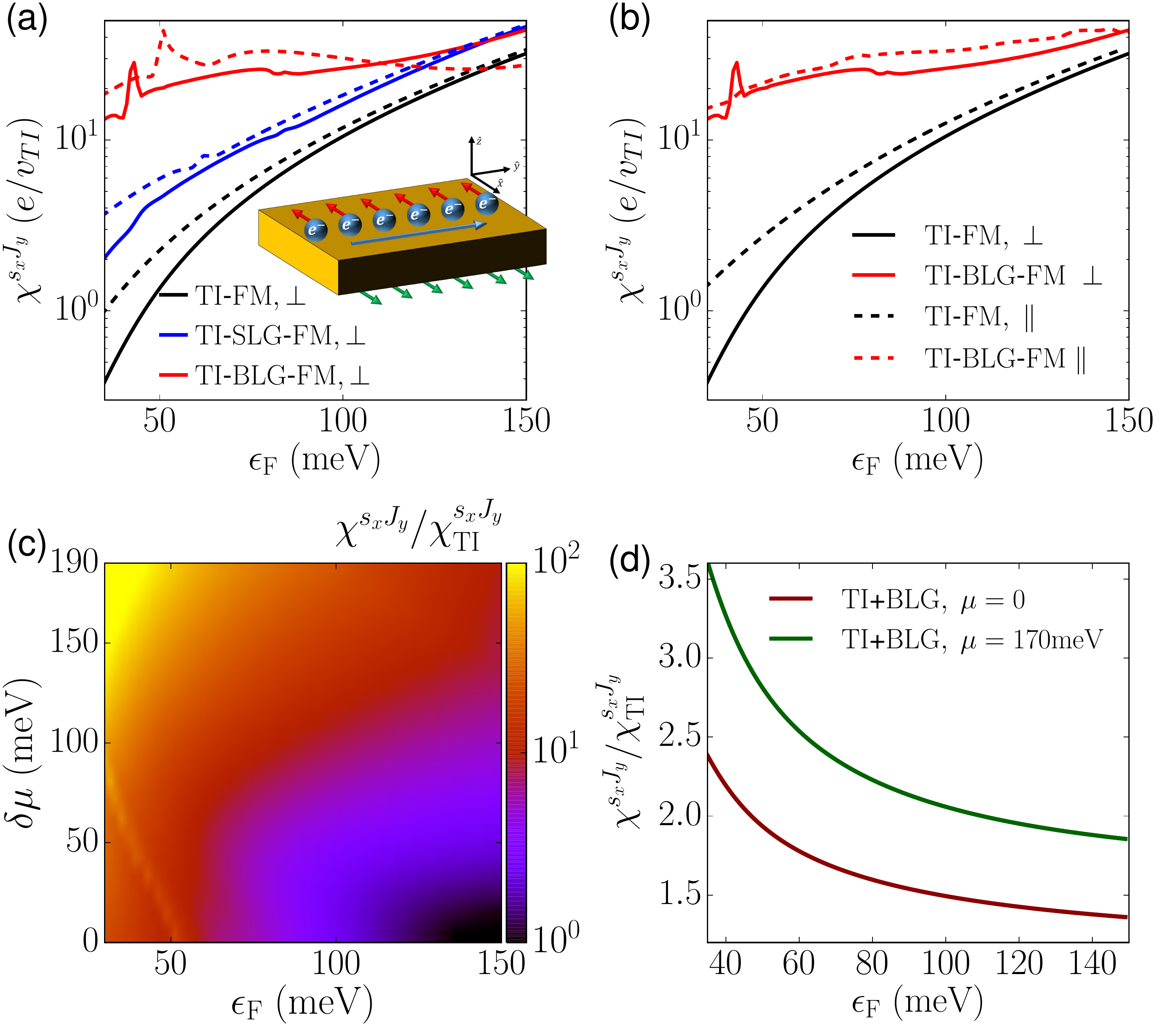}
	\caption{
		(a) $\chi^{s_xJ_y}$ as a  function of $\epsilon_F$ for $\delta\mu=0$, $t=45$~meV, and $\Delta=20$~meV ($\Delta=0$)
		with out-of-plane magnetization $\hat m = \hat z$ ($\perp$), solid (dashed) lines. Inset: sketch showing the spin density accumulation on the top and bottom surface of a TI induced by a current 
		in the $y$ direction.
		(b) $\chi^{s_xJ_y}$ as a function of $\epsilon_F$ for in-plane $\hat m = \hat x$  ($\parallel$) (out-of-plane $\hat m = \hat z$ ($\perp$)) magnetization and $\Delta=20$~meV, dashed (solid) lines. The other parameters are the same as in (a).
		(c) Enhancement of $\chi^{s_xJ_y}$ in a TI-BLG system compared to TI alone as a function of $\epsilon_F$ and $\delta\mu$ for $\Delta=0$.
		(d) $\chi^{s_xJ_y}$ for TI-BLG when $t=0$. In all the panels, the disorder parameters are $n_{\rm imp}=10^{12}$ cm$^{-2}$, and $d=1$~nm. 
		} 
	\label{fig:full_response_vc}
\end{figure}

Figure~\ref{fig:full_response_vc}~(a) shows the dependence of  $\chi^{s_x J_y}$
on $\epsilon_F$ for TI, TI-SLG, and TI-BLG for $t=45$~meV, $\delta \mu=0$ and $\Delta=20$~meV 
with out-of-plane magnetization $\hat m = \hat z$ (solid lines). The dashed lines
corresponds to the case $\Delta=0$. The inset shows a sketch of the system, with
charge flowing in the $y$-direction. The direction of the spin accumultion
on the top and bottom layer is indicated by the arrows on the electrons.
The insertion of a graphene layer strongly 
enhances the current-induced spin density response and therefore the SOT.
Now, we consider in-plane magnetization. In this case, the Fermi surface is not 
isotropic as for out-of-plane magnetization, which makes the computation of scattering 
time, transport times, and the Edelstein effect more challenging. For concreteness, 
we assume the magnetization direction to be $\hat m = \hat x$ ($\parallel$).
Fig.~\ref{fig:full_response_vc}~(b) shows $\chi^{s_xJ_y}$ 
as a function of $\epsilon_F$ for in-plane $\hat m = \hat x$ ($\parallel$) magnetization
and $\Delta=20$~meV, dashed lines. The red lines correspond to a TI-BLG-FM and the black 
lines to a TI-FM heterostructure. We obtained an enhancement as large as 
the one obtained for out-of-plane magnetization $\hat m = \hat z$ ($\perp$), solid lines. 

We find that changes in $\delta\mu$ have a strong impact on $\chi^{s_x J_y}$.
Figure~\ref{fig:full_response_vc}~(c) shows that by increasing $\delta\mu$ the 
enhancement of the SOT can be raised to values as high as 100 in TI-BLG heterostructures
%ER-NN
due to the flattening and consequent increase of the DOS of the TI-like bands
(see Appendix \ref{app:blgbands}). 
%---
%
The results of Fig.~\ref{fig:full_response_vc} show that in TI-SLG and TI-BLG heterostructures
the current-induced SOT can be expected to be much higher than in TI surfaces alone.
They show that for TI-BLG systems there is a large range of values of $\delta\mu$, $\epsilon_F$
for which the enhancement 
of $\chi^{s_x J_y}$ due to the presence of the BLG is consistently close to 10 or larger, 
Fig.~\ref{fig:full_response_vc}~(c).

\begin{figure}[!b]
	\centering
	\includegraphics[width=8.5cm]{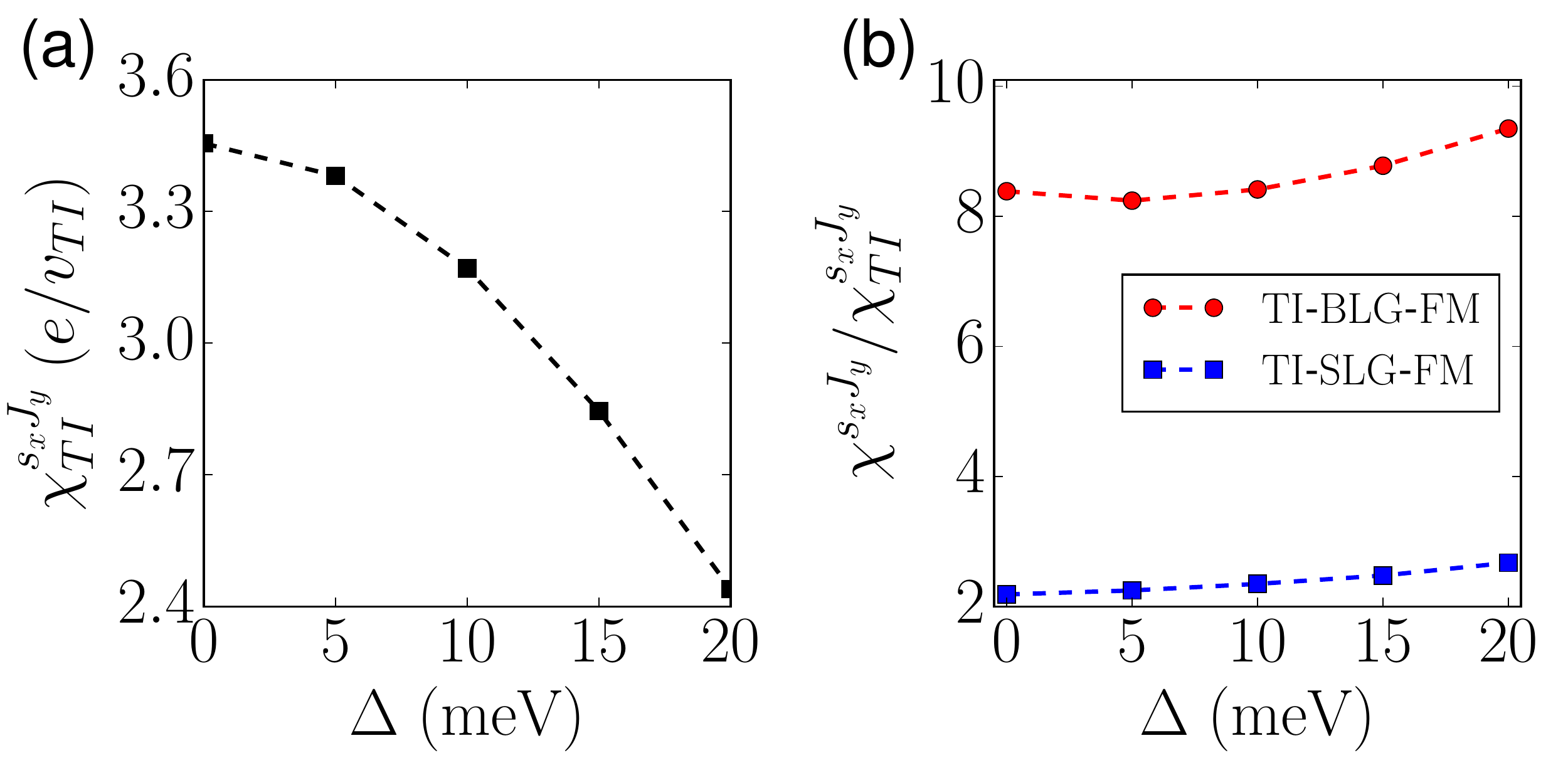}
	\caption{(a) $\chi^{s_x J_y}$ as a function of the exchange interaction $\Delta$ for
		a TI-FM heterostructure at $\epsilon_F = 60$ meV. 
		(b) Ratio $\chi^{s_x J_y}/\chi^{s_x J_y}_{TI}$ of the TI-BLG-FM (red circles) and
		TI-SLG-FM (blue squares) response to the TI-FM response. The magnetization direction
		is out-of-plane, $\hat m = \hat z$ ($\perp$).		
	}
	\label{fig:exchange}
\end{figure}

We also find that the strong enhancement
of $\chi^{s_x J_y}$ is not affected significantly by the value of $\Delta$, 
as shown in Fig. (\ref{fig:exchange}), where we plot $\chi^{s_x J_y}$ as a function of  $\Delta$
at $\epsilon_F = 60$~meV. In Fig. (\ref{fig:exchange})~(a) we plot $\chi^{s_x J_y}$ for
TI-FM, while (\ref{fig:exchange})~(b) shows the response function $\chi^{s_x J_y}$ 
for TI-SLG(BLG)-FM normalized to the response in a TI-FM system.

%
%
%
%ER-NN
%
%In the TI-FM bilayer studied in Ref.~\onlinecite{mellnik2014b} it was shown that 
%$B_{SO}$ (for $I=7.7~{\rm mA}$)~\cite{note2016}
%acting on the FM's magnetization
%was $\sim 3\times 10^{-5} $~T. Given that the presence of a BLG sheet between the TI and the FM conservatively
%enhances $\chi^{s_x J_y}$ by a factor of 10, we estimate that in a TI-BLG-FM system analogous to
%the one studied in  Ref.~\onlinecite{mellnik2014b}, and in the same conditions, $B_{SO}$ could be of the order of
%$\sim 0.3$~mT.
%
%
%Similarly, for a magnetically doped TI the results of Ref.~\onlinecite{fan2016} show that
%$B_{SO}\approx 80$~mT (for $I=4\mu{\rm A}$)~\cite{note2016}.
%%
%Our results therefore show that by placing a BLG sheet on the top surface
%of a magnetically doped TI, in the same condition as in Ref.~\onlinecite{fan2016}, $B_{SO}$ could be as high as 800~mT.
%
%-----
%
In addition, in a TI-graphene heterostructure, by placing the source and drain on the graphene (BLG) and taking into
account the high mobility of graphene (BLG), it is possible to force most of the current to flow
within graphene (BLG) and the TI's surface adjacent to it. Therefore, we can minimize the amount
of spin density accumulation with opposite polarization that a current flowing in the TI's bottom surface generates. 
This fact should further increase the net SOT.

%ER-NN-------
%{\bf XX}
%
The large enhancement of the spin density accumulation in TI-graphene systems is due to two main reasons:
(i) the survival, after hybridization, of TI-like bands well separated from Rashba bands;
(ii) the strong enhancement of the relaxation time $\tau_0$ and transport time $\tau_t$ due to the additional
screening by the graphene layer of the dominant source of disorder.
It is important to notice that the presence of the Rashba bands, see Fig.~1, not only is not
essential for the enhancement of the spin density accumulation but it can be detrimental given that the Rashba bands
give a $\chi^{s_x J_y}$ with opposite sign of the TI-like bands. This fact can be seen at large Fermi energies for 
BLG-TI in Fig.~\ref{fig:full_response_vc}~(a):
for $\epsilon_F\gtrsim 140$~meV the Fermi surface intersect the Rashba bands that by giving
a contribution to $\chi^{s_x J_y}$ opposite to the TI-like bands brings the net SOT of TI-BLG to be slightly lower than the SOT of TI-alone.
Point (ii) explains the fact the $\chi^{s_x J_y}$, at low energies, is always larger in TI-BLG rather than TI-SLG
given that $\tau_0$ and $\tau_t$ are larger in TI-BLG than in TI-SLG.
% due to the fact that BLG at low energies
%has a larger DOS. 
In addition, it explains the fact that even in the limit when there is no hybridization
between the TI and the graphene bands, 
i.e. t=0 due for example to a large twist angle (see Appendix \ref{app:twist}), 
the spin-current correlation function in TI-graphene systems is still larger than in TIs alone
for the experimentally relevant
case where charge impurities are the dominant source of disorder,
as shown in Fig.~\ref{fig:full_response_vc}~(d).

%i.e. t=0 
%due for example to a large twist angle (see Appendix \ref{app:twist}),
%the SOT in TI-graphene systems is still larger than in TIs alone 
%for the experimentally relevant
%case where charge impurities are the dominant source of disorder,
%as shown in Fig.~\ref{fig:full_response_vc}~(d).

\begin{figure}[!h]
	\centering
	\includegraphics[width=8.5cm]{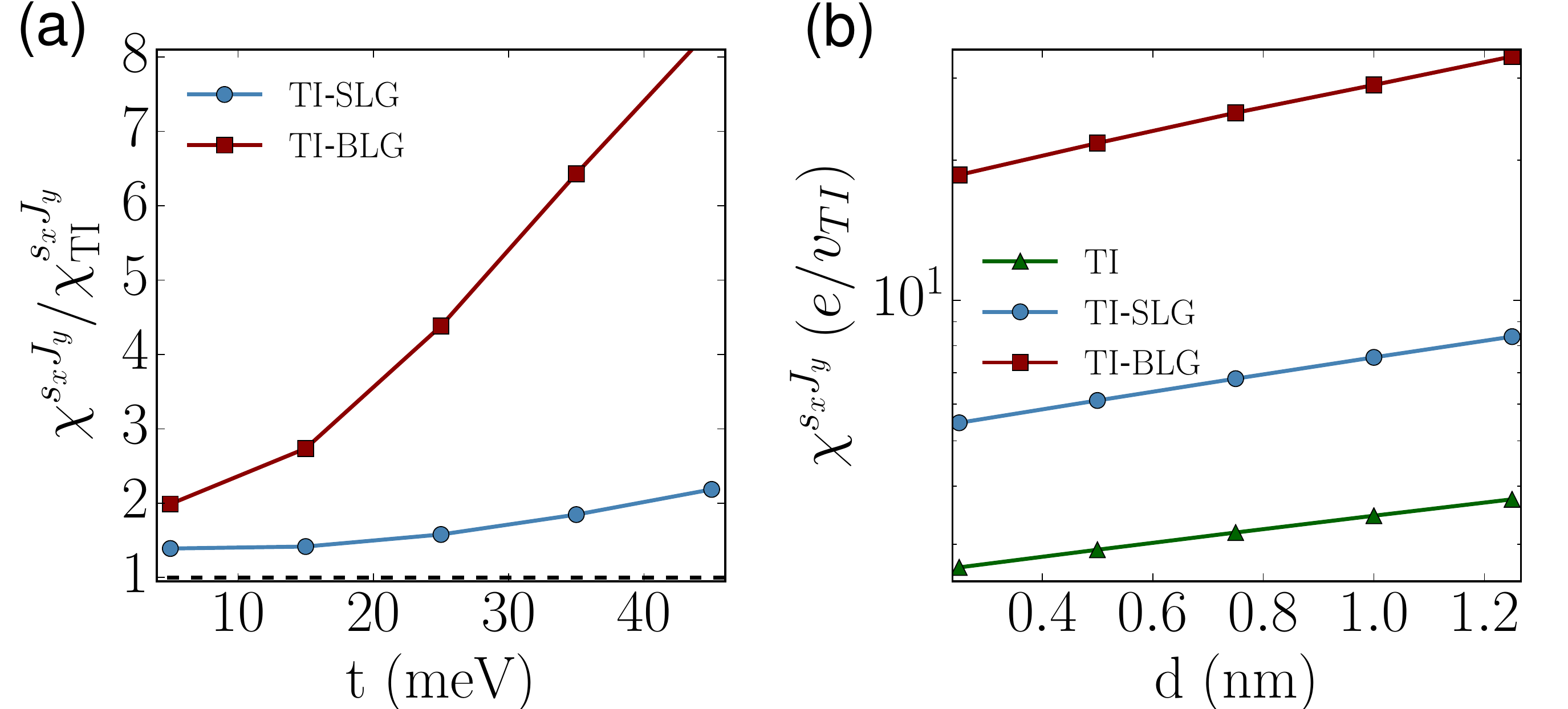}
	\caption{
		(a) $\chi^{s_x J_y}/\chi^{s_x J_y}_{TI}$ as a function of the tunneling amplitude $t$ for
		TI-SLG and TI-BLG heterostructures at $\epsilon_F = 60$ meV and $\delta \mu =0$.
		(b) $\chi^{s_x J_y}$ as a function of the effective distance to the impurities $d$ for
		TI-SLG and TI-BLG heterostructures at $\epsilon_F = 60$ meV and $\delta \mu =0$.
	}
	\label{fig:parameter_space}
\end{figure}

In Fig. \ref{fig:parameter_space}~(a), we show the current-induced spin density accumulation 
response function dependence on the tunneling amplitude $t$, normalized to the TI response.
As $t$ is increased, TI and graphene hybridize more strongly, leading to a larger SOT. However,
even at vanishing tunneling, an enhancement is still present. 

In Fig. \ref{fig:parameter_space}~(b), we plot $\chi^{s_x J_y}$ as a 
function of the effective distance from the TI surface to the effective
layer of impurities $d$. The further away the impurities are located, 
the weaker the disorder, and therefore the larger the expected SOT. 

%
%-------------
%
%For this reason we also expect that even when the graphenic layer and the TI are in an incommensurate
%stacking configuration, the presence of the graphenic layer, by screening the dominant source of disorder,
%will lead to an enhancement of the SOT~\cite{sm}.

%
%

\begin{figure}[!t]
	\includegraphics[width=8.5cm]{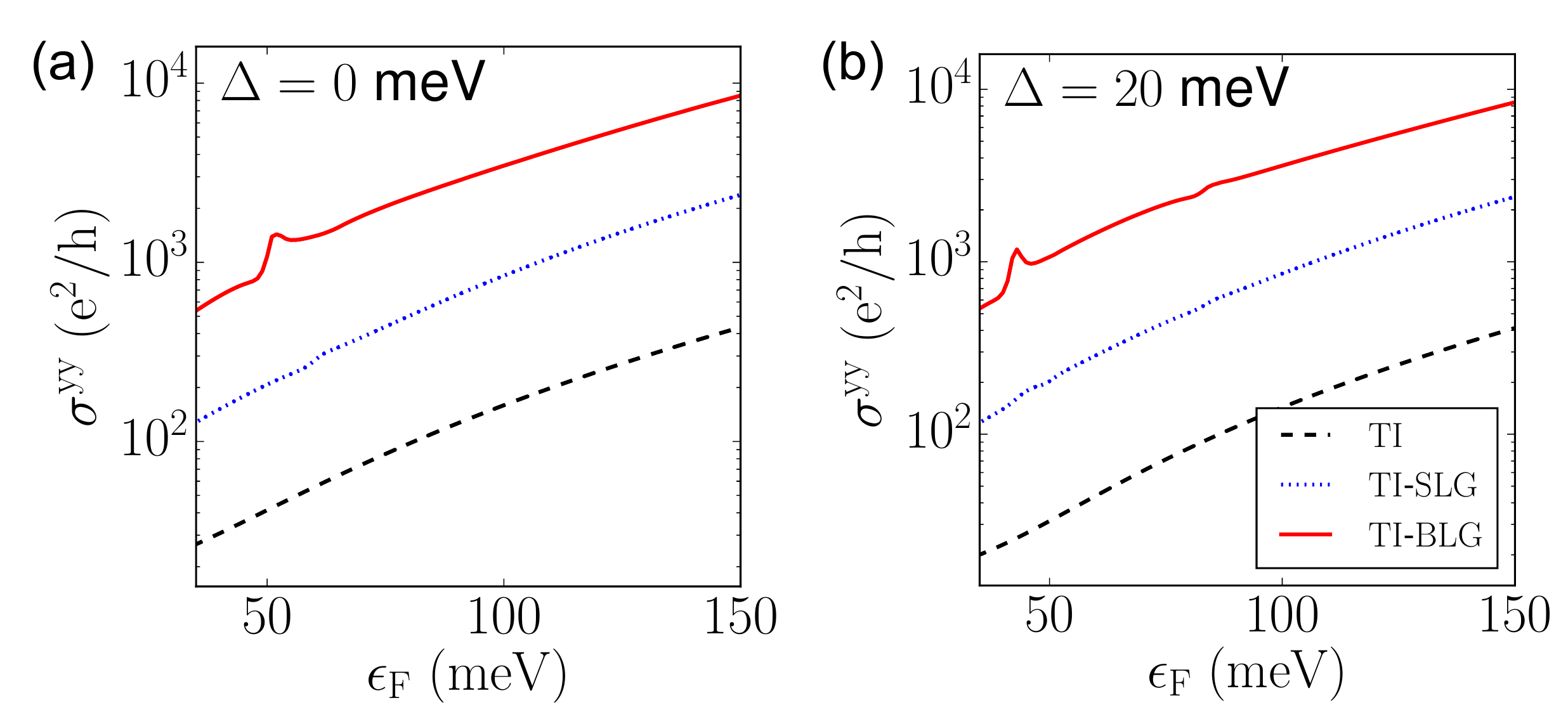}
	\caption{
		$\sigma^{yy}(\epsilon_F)$, for TI (dashed line), TI-SLG (dotted line), and TI-BLG (solid line)
		for $\Delta=0$, (a), and $\Delta=20$~meV (b) with out-of-plane magnetization, $\hat m = \hat z$ ($\perp$).
		$t=45$~meV, $\delta \mu=0$, and $n_{\rm imp}=10^{12}$ cm$^{-2}$.
	}
	\label{fig:sigma}
\end{figure}

%
To estimate the efficiency of the current-induced SOT in TI-graphene heterostructures,
we calculate the associated dc longitudinal conductivity $\sigma^{ii}$ for the same parameters.
In the linear-response, long-wavelength, regime we have 
\begin{align} 
\sigma^{ii} & \approx   \frac{e^2}{2 \pi \Omega} \Re \sum_{\kk, a}  v^i_{aa}(\kk)\tilde{v}^i_{aa}(\kk) G^A_{\kk a}G^R_{\kk a}\;.
\label{eqn:conductivity}
\end{align}
Fig.~\ref{fig:sigma}~(a)
shows $\sigma^{yy}$ for TI, TI-SLG, and TI-BLG as a 
function of $\epsilon_F$ in the limit $\Delta=0$.  We see that the presence of a graphene layer enhances 
the conductivity of the system by an order of magnitude or more.
%For TI-BLG $\sigma^{ii}$, the enhancement is a about a factor
%50 larger than in the TI alone.
%
%
Fig.~\ref{fig:sigma}~(b) shows that
the exchange term $H_{ex}$ 
does not affect $\sigma^{yy}$ significantly.
The results shown in Fig.~\ref{fig:sigma}~(b) imply that in TI-graphene heterostroctures
not only the current-induced SOT can be much larger than in TIs alone, but also that the
generation of the SOT is much less dissipative. 
For example, for an applied electric fields of the order of 0.1 $V/\mu$m, 
we can reach a conservative spin density accumulation
$\delta s^x \approx 5 \times 10^7 \hbar \;\mbox{cm}^{-2}$.
For typical carrier density in graphene ($n \approx 10^{11} \mbox{cm}^{-2}$), we have 
$\delta s^x/n = 5\times 10^{-4} \hbar$.

%
%ER-NN
%{\bf XX}
\section{Conclusions}
\label{sec:conclusions}
In conclusion, we have shown that
in magnetic TI-graphene heterostructures the
non-equilibrium uniform spin density accumulation induced by a charge current can be 10-100 times
higher than in TIs alone giving rise to a giant Edelstein effect.
The reasons for these enhancements are 
(i)   the additional screening by the graphene layer of the dominant source of disorder;
(ii)  the fact that graphene and the TI's surface are almost commensurate making possible a strong hybridization of the TI's and graphene's states;
(iii) the fact that the spin structure of the hybridized bands has a spin structure very similar to the one of the original TI's band
      for a large range of dopings;
(iv)  the fact that graphene is the ultimate 2D system, only one-atom thick.
%, and therefore does not ``short-circuit'' the TI's surface.
%
%We also find that in TI-graphene systems the 
%charge conductivity is 10-50 times larger than in TI surfaces.
%Therefore we predict very large SOTs with low dissipation in these systems.
%
%
These facts and our results suggest the TI-graphene systems are
very good candidates to realize all-electric efficient magnetization switching.

\section*{ACKNOWLEDGEMENTS}
We acknowledge helpful discussions with Yong Chen and Saroj Dash. 
MRV and ER acknowledge support from NSF Grant No. DMR-1455233 and ONR Grant No. N00014-16-1-3158. ER also acknowledges support from ARO Grant No W911NF-16-1-0387, and the United States-Israel Binational Science Foundation, Jerusalem, Israel. In addition, MRV and ER thank the hospitality of the Spin Phenomena Interdisciplinary Center (SPICE), where this project was initiated.
JS and GS acknowledge the support by Alexander von Humboldt Foundation, the ERC Synergy Grant SC2 (No. 610115), and the Transregional Collaborative Research Center (SFB/TRR) 173 SPIN+X.

\appendix
%%%%%%%%%%%%%%%%%%%%%%%%%%%%%%%%%%%%%%%%%%%%%%%%%%%%%%%%%%%%
\section{TI-BLG BAND STRUCTURE}
\label{app:blgbands}
As long as the interlayer tunneling $t_{BLG}$ between the carbon
atoms in bilayer graphene is much larger than the expected
tunneling $t$ between the TI's surface and the graphenic layer
any difference between the tunneling strength between 
the carbon layers forming BLG and the TI will give very negligible effects.
Considering that in bilayer graphene the interlayer tunneling is 350~meV,
and the fact that for the TI-graphene tunneling $t$ we only consider
values smaller than 45~meV for all our results is $t\ll t_{BLG}$.
In this limit, at low energies ($\lesssim 350$~meV), BLG can be
treated as 2D system with the effective Hamiltonian $H^{BLG}$
presented in the main text.

\begin{figure}[!b]
	\centering
	\includegraphics[width=7.0cm]{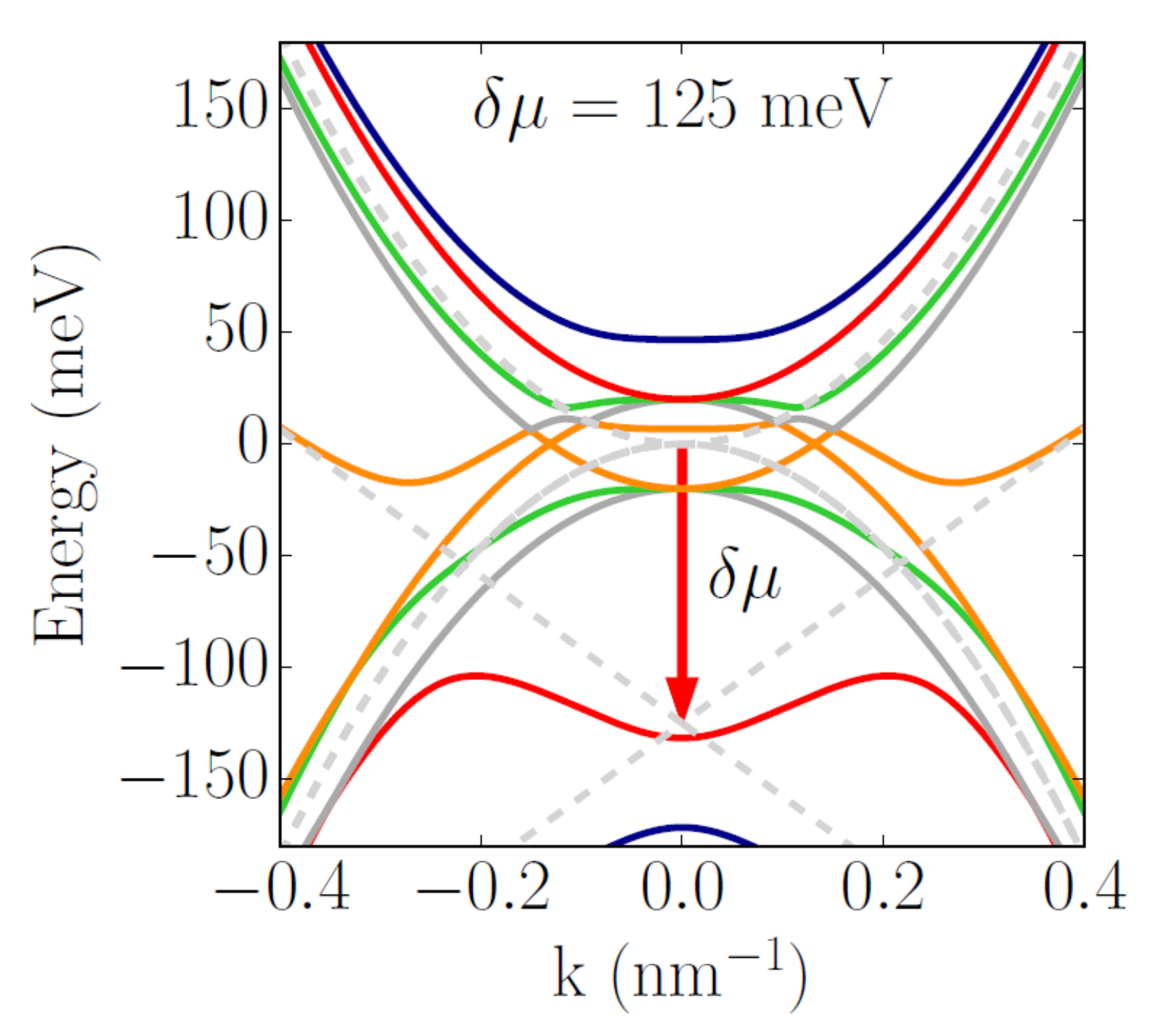}
	\caption{TI-BLG band structure for $\delta \mu = 125 $ meV, and tunneling amplitude
		$t=45$~ meV.}
	\label{fig:bands_1}
\end{figure}

Fig.~1~(e) in the main text shows the bands of a TI-BLG systems
for which the exchange field $\Delta=20$~meV and $\delta\mu=0$.
Fig~3 shows that the strongest enhancement of the SOT happens
for TI-BLG systems when $\delta\mu\neq 0$. It is therefore
interesting to see how the low-energy bands of TI-BLG 
are affected by a nonzero value of $\delta\mu$.
Fig.~\ref{fig:bands_1} shows the band structure of TI-BLG
for the case when $\delta \mu = 125 $ meV in the absence of any exchange field.
We see that one of TI-like bands (shown in orange) becomes much flatter:
the high density of states of this band explains the high values of SOT
for TI-graphene systems when $\delta\mu\neq 0$.\\

\section{INVERSE SPIN GALVANIC EFFECT IN TWISTED TI-GRAPHENE HETEROSTRUCTURES}
\label{app:twist}
It can expected that even when the stacking of the graphenic layer and the TI's surface is incommensurate,
the screening of the charge impurities by the graphenic layer will lead to a strong enhancement of $\tau_0$
and $\tau_t$ and therefore of the SOT. The accurate treatment of the realistic case in which the main
source of disorder are charge impurities for incommensurate stackings
requires the calculation of
the dielectric constant for incommensurate structures, a task that is beyond the scope of the present work.
For this reason, to exemplify how the presence of a small twist angle $\theta$ between the graphenic layer and the TI surface, 
giving rise to an incommensurate stacking, affects the calculation of the SOT, we consider 
a very simple
model for the effect of the disorder: we simply assume that the disorder gives rise to a constant quasiparticle broadening.

\begin{figure}[!th]
	\centering
	\includegraphics[width=8.5cm]{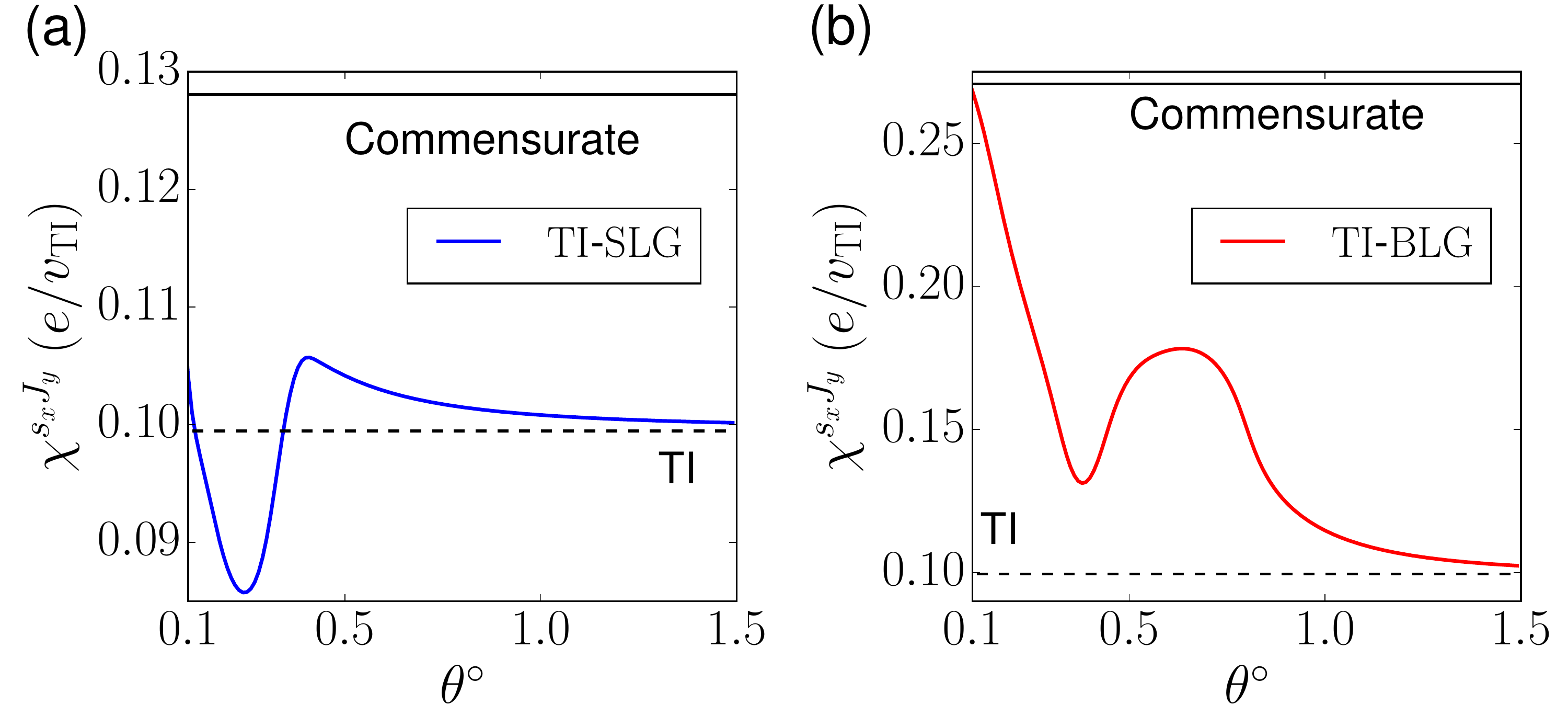}
	\caption{(a) $\chi^{s_x J_y}$ as a function of twist angle $\theta$ for TI-SLG.
		(b) Same as (a) for TI-BLG.}
	\label{fig:twisted}
\end{figure}

Let $|\mathbf{q}|\equiv q=2K_{D}\sin(\theta/2)$, where $K_{D}$ is the magnitude
of the graphene $K$ point. The dimensionless
parameter $\gamma\equiv\frac{t'}{\hbar v_{TI}q}$, where $t'=t/3$, measures the strength of the coupling between
the graphenic layer and the TI. For $\gamma<1$ we can obtain the electronic structure using the weak coupling theory
for twisted systems~\cite{dossantos2007,bistritzer2011,mele2012} that
for the case of a TI-graphene heterostructures we presented in Ref.~\cite{jzhang2014}.
After obtaining the electronic structure in the regime $\gamma<1$, we can obtain $\chi^{s_x J_y}$.
To understand how the response between the commensurate and the incommensurate regimes differ,
we have calculated  $\chi^{s_x J_y}$ assuming a constant quasiparticle broadening
$1/(2\tau_0)=2$~meV, with $t'=15$~meV, $\delta \mu=0$, and $\epsilon_F=10$~meV for a range of 
values of $\theta$ for which the weak coupling
theory is valid. The dependence of $\chi^{s_x J_y}$, per valley, as a function
of $\theta$ for TI-SLG and TI-BLG is shown in Fig.~\ref{fig:twisted}.
As to be expected, the results show that in the incommensurate case the response is
smaller than in the commensurate case. However, they also show, in particular for the case in which the 
graphenic layer is BLG, that a TI-graphene heterostructure is expected to have stronger $\chi^{s_x J_y}$, 
and therefore a stronger inverse spin-galvanic effect, even in the incommensurate regime
and for the case in which the disorder is modeled very simply.
\clearpage
%%%%%%%%%%%%%%%%%%%%%%%%%%%%%%%%%%%%%%%%%%%%%%%%%%%%%%%%%%%%%%%%%%%%%%%%
%\bibliographystyle{apsrev4-1}
%\bibliography{spin,bib2,spin2}

\begin{thebibliography}{65}
\bibitem [{not({\natexlab{b}})}]{note2016b}%
	\BibitemOpen
	\href@noop {} {\bibinfo {title} {{\rm Strictly speaking the value
					$\kappa_0=1$ is only valid for the case of a magnetically doped TI. However,
					given the large static dielectric constant of the TI, the error made by
					approximating the FM's dielectric constant by the vacuum's is negligible. For
					EuS, the suggested insulating FM, $\kappa_0\approx 10$ giving an effective
					average dielectric constant $\kappa=55$, instead of the used $\kappa=50$.
			}}}
\end{thebibliography}


\end{document}